\documentclass[review,3p,times,sort&compress,letterpaper]{elsarticle}              

\usepackage{amsmath}
\usepackage{amssymb}
\usepackage{graphicx}
\usepackage{epsfig}
\usepackage{color}
\usepackage{adjustbox}
\usepackage{ctable}
\usepackage{float}
\usepackage{hyperref}
\usepackage{xspace}
\usepackage{colortbl}       

\usepackage{lmodern}
\usepackage{mathtools}
\usepackage{mathpazo}
\usepackage[squaren,Gray]{SIunits}

\usepackage[utf8]{inputenc}
\usepackage[T1]{fontenc}

\let\OldTexttrademark\texttrademark
\renewcommand{\texttrademark}{\OldTexttrademark\xspace}

\journal{Materials \& Design}

\begin{document}

\begin{frontmatter}


\title{Accelerated Fatigue Strength Prediction via Additive Manufactured Functionally Graded Materials and High-Throughput Plasticity Quantification}

\author[ca,uiuc]{C. Bean}\ead{cmbean2@illinois.edu}
\author[uiuc]{M. Calvat}
\author[uiuc]{Y. Nie}
\author[uiuc]{R.L Black}
\author[uiuc]{N. Velisavljevic}
\author[uiuc]{D. Anjaria}
\author[ca,uiuc]{M.A. Charpagne}
\author[ca,uiuc]{J. C. Stinville}\ead{jcstinv@illinois.edu}

\cortext[ca]{corresponding author}

\address[uiuc]{University of Illinois Urbana-Champaign, Urbana, USA}

\begin{abstract}
Recent improvements in additive manufacturing and high-throughput material synthesis have enabled the discovery of novel metallic materials for extreme environments. However, high-fidelity testing of advanced mechanical properties such as fatigue strength, has often been the most time-consuming and resource-intensive step of material discovery, thereby slowing down the adoption of novel materials. This work presents a new method for rapid characterization of the fatigue properties of many compositions while only testing a single specimen. The approach utilizes high-resolution digital image correlation along with a computer vision model to extract  the relationship between localized plastic deformation events and associated mechanical properties. The approach is initially validated on an additive manufactured 316L dataset, then applied to a functionally graded additive manufactured specimen with a composition gradient across the gauge length. This allows for the characterization of multiple compositions, orders of magnitude faster than traditional methods. 
\end{abstract}


\end{frontmatter}

\noindent \textbf{Highlights}

\begin{itemize}
  \item High-throughput fatigue strength estimation of 5 chemical compositions from a single mechanical test
  \item 20-fold acceleration in fatigue strength prediction
  \item Functionally graded materials for rapid investigation of fatigue strength
\end{itemize}

\section{Introduction}

\begin{figure}
    \centering
    \includegraphics[width=1\textwidth]{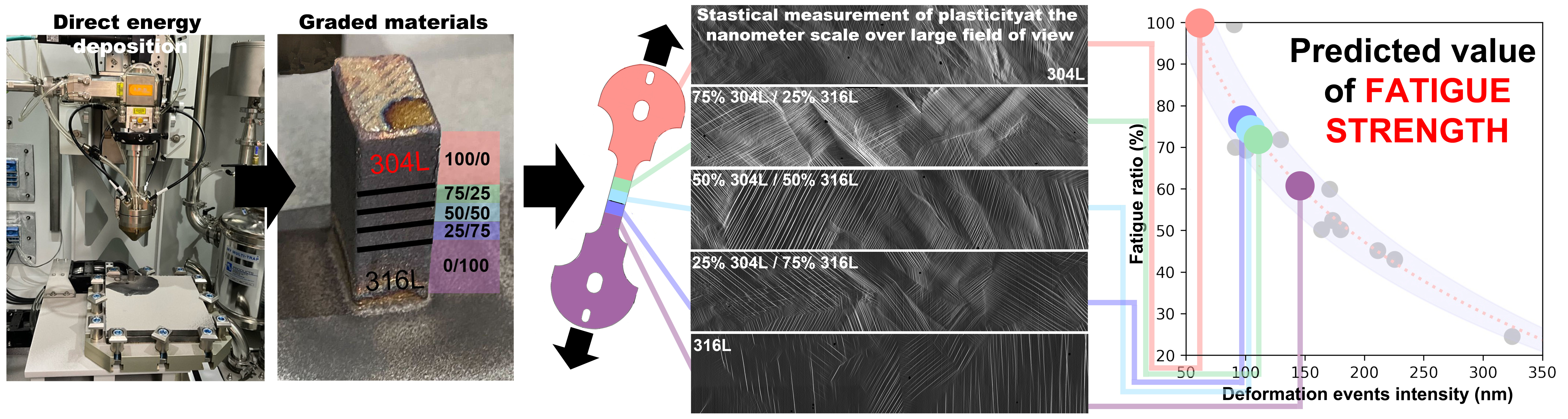}
    \label{GraphicalAbstract}
\end{figure}

Significant advances in high-throughput material synthesis and characterization complemented by computational and machine-learning techniques have brought the longstanding goal of accelerating material discovery within grasp. With these advancements, materials with favorable chemical composition, crystal structure, and microstructure can be swiftly identified \cite{Rao2022, WEN2019109,Ramprasad2017, PhysRevB.89.094104, Zhu2021, Farrag2024, LIU2022224}. However, characterization of their advanced mechanical properties such as fatigue strength, creep resistance, fracture toughness, wear, and impact resistance remain time-intensive and an expensive aspect of materials down-selection and validation \cite{GIANOLA2023101090,Farrag2024,MIRACLE2024101188}. Advanced mechanical characterization considerably hampers the materials development cycle, yet it is undeniably a vital step in evaluating the performance improvements potentially offered by new materials. While advanced testing can be reduced through down-selection based on previous knowledge and rapid screening techniques, it cannot be completely eradicated due to the ultimate need to measure design-relevant properties. Macroscopic testing of higher fidelity (including creep, fatigue, fracture, and impact) is time-consuming and remains the sole method to obtain these advanced mechanical properties. Recently developed automated and parallel testing methods, which consist of testing multiple specimens in parallel or in series using robots, hold promising potential for speeding up material properties evaluation \cite{https://doi.org/10.1002/smtd.202201591}. Simultaneously, full-field measurements focusing on macroscopically important quantities like stress and strain across a wide area are employed to obtain the material's response under various loading conditions in just one test \cite{annurev:/content/journals/10.1146/annurev-matsci-080619-022100}. However, using either full-field measurements, automation, or parallel testing methods, the duration of each test for a given specimen remains identical to traditional advanced mechanical testing and the gain in time is therefore limited. More recently, alternative high-throughput methods have employed inverse analysis on local measurements to predict macroscopic properties. These methods are particularly interesting as they offer a swift screening of material properties. They are all based on inverse analysis of a material's behavior at a microstructure-scale to predict macroscopic properties. The most common method involves the extraction of monotonic properties through local indentation measurements \cite{doi:10.1073/pnas.1922210117, CLAUSNER201511}. Regarding advanced mechanical properties such as creep, fatigue, or fracture, identifying the controlling features that can predict advanced macroscopic properties remains largely elusive, with the recent exception of fatigue strength \cite{science2022}.
Links between fatigue strength and the characteristics of nanometer-scale deformation events that develop during an initial fully (reversed) loading cycle exist \cite{science2022}. It was observed that the statistical and quantitative investigation of the intensity of nanoscale plastic deformation events that develop during a representative loading, i.e. tensile (and compressive) test at the material yield strength (and reverse loading to no residual macroscopic plastic strain) can inform the fatigue strength of metallic materials \cite{science2022}. These measurements can be acquired in just a few hours \cite{Black2023}. Therefore, the relationship between plastic deformation event characteristics during simple elementary loading and long-term properties may allow for the rapid evaluation of advanced mechanical properties. In this study, we exploit the relationship between fatigue strength and intensity of nano-scale plastic deformation events along with automated and high-throughput measurements of plastic deformation events, and computer vision-assisted segmentation of these plastic deformation events to rapidly predict the fatigue strength of materials. 

We leverage emergent additive manufacturing (AM) technologies and their ability to synthesize metallic materials with various compositions and microstructures in a single step \cite{ZHANG2019138209,BOBBIO2017133}. Of interest, AM allows the fabrication of functionally graded materials (FGMs) with spatially controlled chemical composition \cite{ZHANG2019138209,BOBBIO2017133,NIE2023115714} and/or microstructure \cite{POPOVICH2017441,ZHANG2019138209} variations in a single specimen. While the presence of multiple materials per specimen would a priori challenge our ability to characterize advanced properties even further, it simultaneously opens opportunities to collect and characterize multiple microstructures, and/or chemical compositions in a single mechanical test. 

Leveraging both the identified relationship between the characteristics of plastic deformation events and fatigue strength \cite{science2022}, and our capability to synthesize FGMs via AM, we report here an unparalleled advancement in high-throughput measurements to estimate the fatigue strength of multiple metallic materials produced by AM. The present method is explained and validated on an AM stainless steel FGM with a linear gradient in nickel-to-chrome ratio.

\section{A High-Throughput Approach}

The current high-throughput method encompasses three key phases: (i) creating a specimen with graded chemical composition/microstructure; (ii) measuring the characteristics of plastic deformation events during an elemental loading step for each region in the graded specimen; (iii) extracting the most intense plastic deformation events for each composition/microstructure in the graded specimen and estimation of fatigue strength for multiple compositions and AM microstructures.

\subsection{Functionally graded material synthesis through additive manufacturing}

\begin{figure}
    \centering
    \includegraphics[width=1\textwidth]{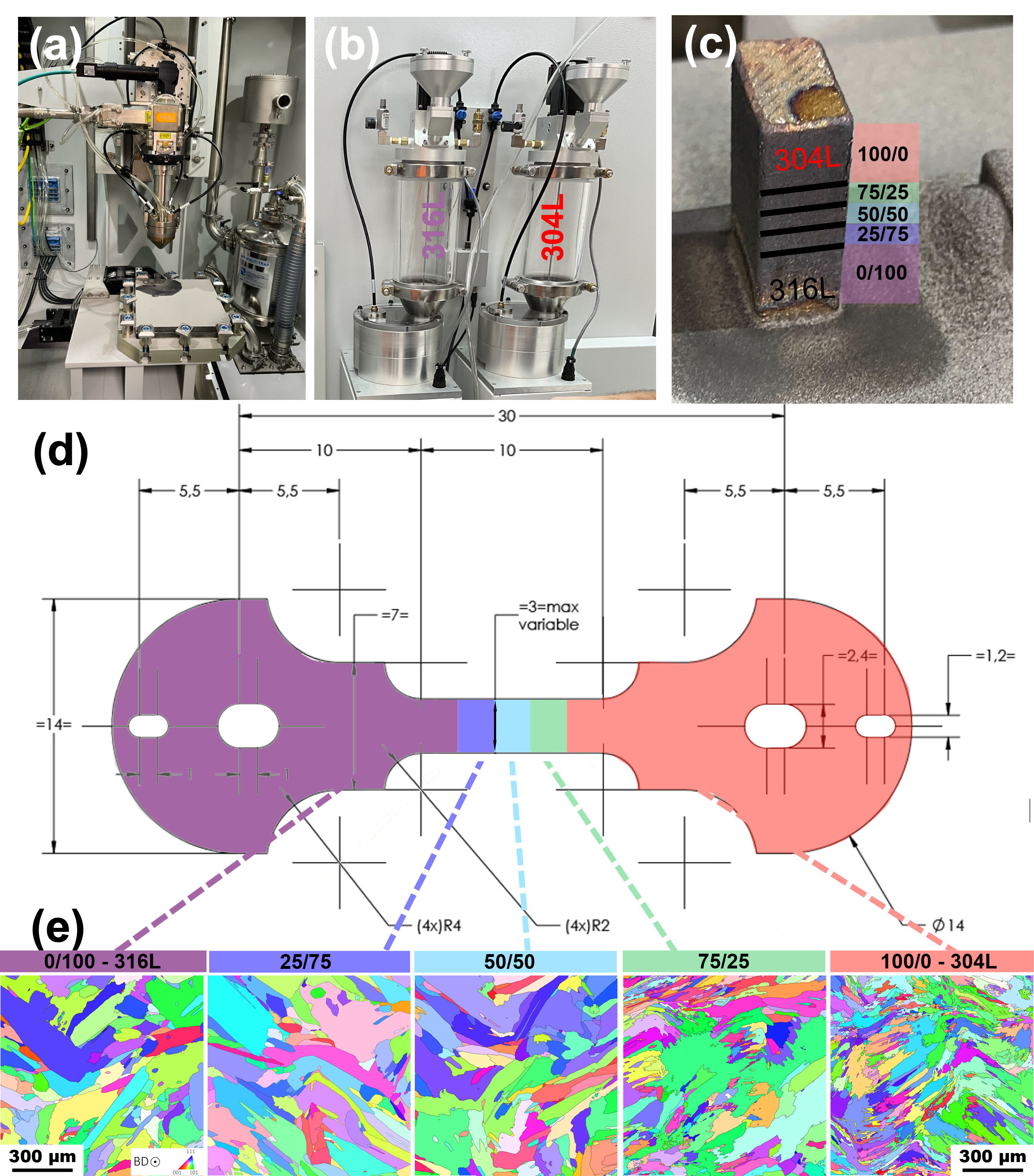}
    \caption{\textbf{Functionally graded material (FGM) specimen and microstructure} (a) The Formalloy L2 direct energy deposition system used to additively manufacture the graded block. (b) The two powder feeders used in the DED systems to produce the graded block. (c) As-printed graded block prior to extraction of the tensile specimen. The colors highlight the 304L/316L powder ratios used during deposition. (d) Tensile specimen geometry of the graded material used for mechanical testing. Dimensions are in millimeters. The thickness of the specimens was 2 mm. (e) EBSD maps of representative regions of interest showing the five materials within the FGM. All EBSD maps are coded in IPF colors with respect to the build direction, horizontal to the surface in (e).}
    \label{FGM}
\end{figure}

Gas atomized 304L and 316L powders with diameter in the 45 $\mu m$ - 105 $\mu m$ range and respective composition (in wt.\%) Fe-18.62\% Cr-9.52\% Ni-1.3\% Mn-0.02\% Cu-0.75\% Si-0.018\% C-0.02\% O-0.01\% P-0.004\% S-0.07\% N and Fe-17.6\% Cr-12.6\% Ni-0.89\% Mn-2.43\% Mo-0.67\% Si-0.019\% C-0.02\% O-0.007\% P-0.004\% S-0.09\% N, were purchased from Carpenter Technologies. N, O, and C were measured by LECO analysis and other elements via inductively coupled plasma optical emission spectroscopy. A 316L to 304L FGM  was printed using a Formalloy L2 direct energy deposition (DED) presented in Figure \ref{FGM}(a), equipped with a 1 kW Nd:YAG laser (wavelength 1070 nm), a Gaussian energy profile, and two powder feeders shown in Figure \ref{FGM}(b). Argon was used as both shielding and carrier gas. A laser power of 500 W, a scan speed of 1100 mm/s, and a scan rotation of 90$\degree$ between adjacent layers were used. The bottom of the block was printed using the 316L alloy. A sequential change from 0/100, 25/75, 50/50, 75/25 to 100/0 304L/316L volume ratio was used over a 10-millimeter height, programmed through G-code. Inter-material regions were also found and analyzed but left out of the analysis due non-uniform chemical composition. The top of the block was printed using the 304L feeder only, as shown in Figure \ref{FGM}(c). The built block was removed from the base plate using a band saw. A flat tensile dogbone-shaped specimen with a gauge section of 1$ \times $3 mm$^2$ and thickness of 2 mm was cut by wire electrical discharge machining, mechanically mirror polished using abrasive papers and diamond suspensions, and chemo-mechanically polished using a suspension of 0.04 $\mu m$ colloidal silica particles. Conventional electron backscatter diffraction (EBSD) measurements were performed at 70$^{\circ}$ tilt using a step size of 1 micron, square grid collection, 200 frames per second acquisition, an accelerating voltage of \unit{30}{kV} and a current of \unit{2.6}{\nano\ampere} before deformation, and collected on the entire gauge length of the specimen. EBSD was performed on a Thermofischer Scios2 Dual-Beam SEM using a Hikari Super EBSD detector. To validate the proposed approach, reference plain blocks of 316L and 304L were printed and prepared using the same AM parameters as the FGM. All AM blocks show a density that exceeds 99.9\%. Microstructural parameters were used to detail each material: grain sizes (equivalent diameter) were estimated from EBSD using a 5$^\circ$ threshold; the grain reference orientation deviation (GROD) represents the angular deviation of the orientation of each point within a grain from its average orientation; mean serration represents the average serration of grain boundaries and was calculated using the approach described in the method in Ref. \cite{NIE2024120035}; the twin ratio is obtained from the ratio between twin boundary ($\Sigma 3$) and total boundary number.

\subsection{Quantitative measurements of plastic deformation events}

\subsubsection{High-Resolution Digital Image Correlation}

\begin{figure}
    \centering
    \includegraphics[width=1\textwidth]{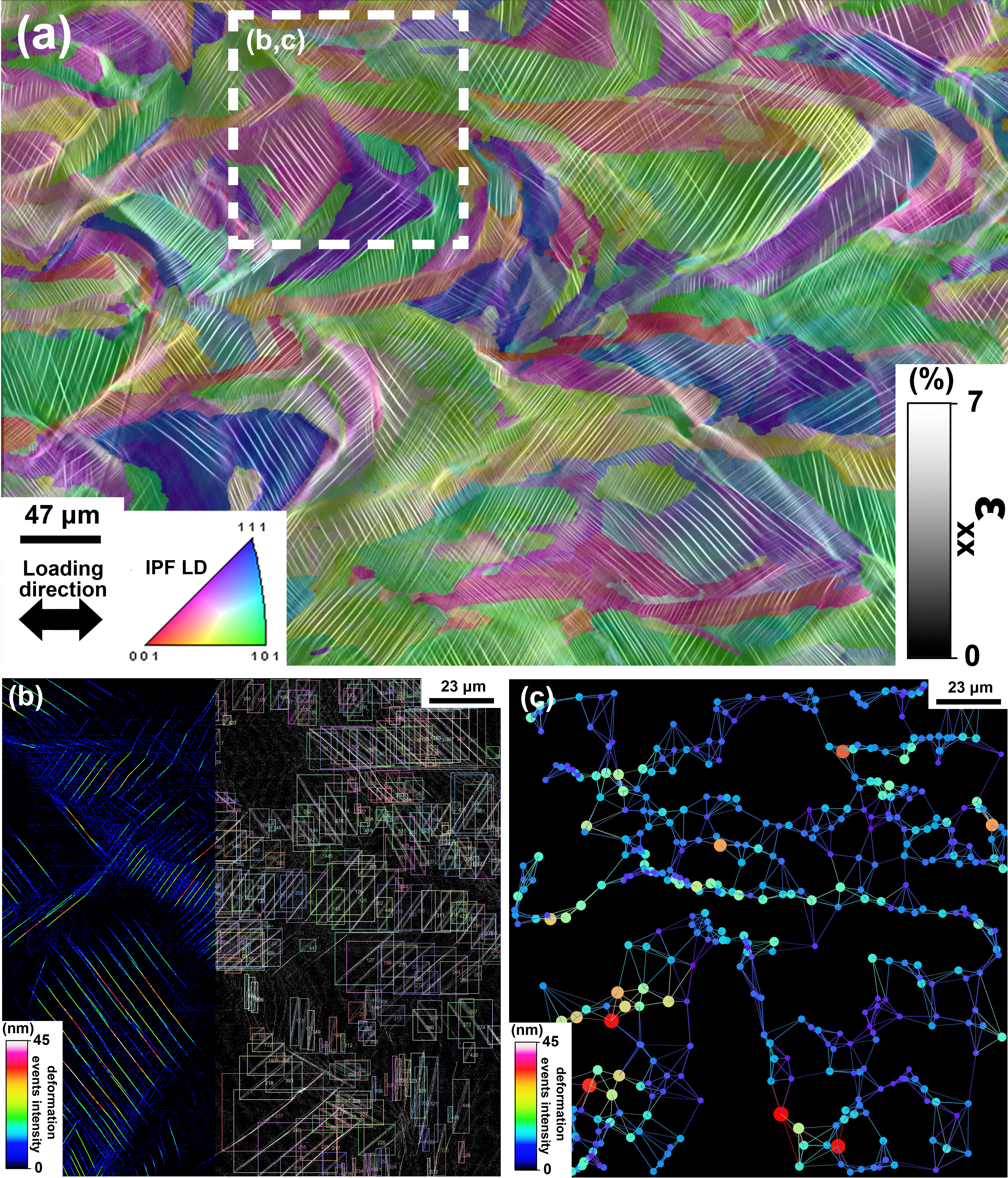}
    \caption{\textbf{Graph representation of plastic deformation events} (a) Strain map ($\epsilon_{xx}$) after deformation of a AM 316L material at 0.72\% macroscopic plastic deformation. This map is superimposed with the associated inverse pole figure map along the loading direction. (b, left) The intensity map of the plastic deformation events within the area marked by the dashed white box in (a). (b, right) Visualization of boxes and masks obtained from a computer vision algorithm created to identify and segment plastic deformation events. (c) Graphical network representation of the plastic deformation events within the area outlined by the dashed white box in (a). Here, each event's size and color denote its maximum intensity along the event, while the connecting lines between events illustrate their connectivity.}
    \label{CV_HRDIC_small}
\end{figure}

Before deformation, a speckle pattern consisting of \unit{60}{\nano\meter} silver nanoparticles was formed on the surface of the sample, following the procedure developed by Montgomery \textit{et al.}~\cite{Montgomery2019}. High-resolution digital image correlation (HR-DIC) was performed using the software XCorrel \cite{Valle2015} during monotonic deformation. HR-DIC images were captured before and after loading across the entire gauge length. The SEM images (6144~px $\times$ 4096~px) were divided into subsets of $31 \times 31$ pixels ($\unit{700}{\nano\meter} \times \unit{700}{\nano\meter}$) with an overlap of 28 pixels between each subset, i.e., step size of 3 pixels ($\unit{67}{\nano\meter}$). A monotonic tensile test was conducted at a quasi-static strain rate between $10^{-5}$~s$^{-1}$ to $10^{-4}$~s$^{-1}$ using a NewTec Scientific$\texttrademark$ MT1000 5~kN tensile stage. Interruptions were performed at macroscopic plastic strains of 0.102\%, 0.218\%, 0.460\%, 1.01\%, 1.65\%, 2.36\%, 3.29\%. Each time the specimen was unloaded, an HR-DIC measurement was performed. Heaviside-based digital image correlation (Heaviside-DIC) was used to address the issue of discontinuities present within displacement fields \cite{Valle2015, BOURDIN2018307, BERGSMO2022104785, BEAN2022103436} due to the occurrence of discrete metallic deformation processes. The Heaviside-DIC method is utilized to extract quantitative measurements of the intensity of discontinuous displacements, i.e. displacement induced along plastic deformation events, across large areas. It employs a localized approach that provides additional analysis within the DIC subsets \cite{Valle2015}. Discontinuities at the subset level are described by Heaviside functions, enabling one part of the subset to shift in relation to others by an amount equivalent to the physical amplitude of the local shearing induced by a plastic deformation event. When mechanical loading occurs, plastic deformation events appear on the free surface of specimens, each associated with a local surface step and in-plane displacement \cite{Mughrabi2009}. The Heaviside-DIC method quantifies the magnitude of the in-plane displacement generated along the deformation event and provides a value in nanometers. An example is provided in Figure \ref{CV_HRDIC_small} for a reduced region in an AM 316L deformed at a macroscopic plastic strain of 0.72\%. The $\epsilon_{xx}$ strain maps display the local strain along the loading direction, which are provided in Figure \ref{CV_HRDIC_small}(a) overlapped with the AM microstructure obtained by EBSD. The bands of concentrated strain are evidence of deformation events, here dislocation slip. The intensity associated with the deformation events is provided in Figure \ref{CV_HRDIC_small}(b, left) for the reduced region of interest depicted by a dashed white box in Figure \ref{CV_HRDIC_small}(a). The intensity of the plastic deformation events is given in nanometers and represents the in-plane displacement induced by the shearing generated by the emergence of the dislocations at the surface of the specimen along a crystallographic plane.

\subsubsection{Graph representations of plastic deformation events from HR-DIC measurements}

\begin{figure}
    \centering
    \includegraphics[width=1\textwidth]{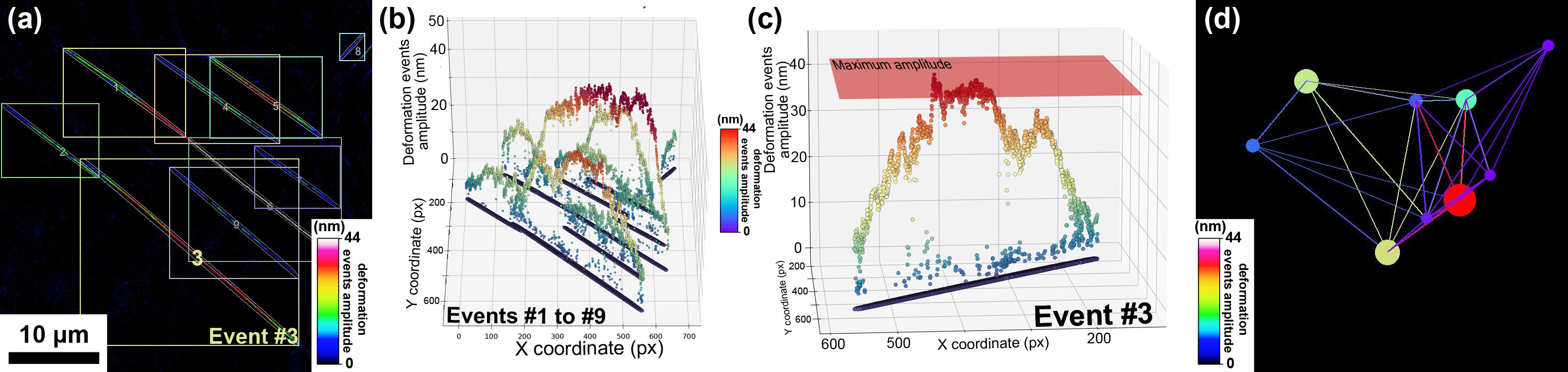}
    \caption{\textbf{Extraction of plastic deformation events characteristics, here maximum intensity} (a) a small region of interest and associated plastic deformation events intensity map. The boxes and masks are obtained from the computer vision model. (b) 3D representation of the pixels extracted from each event using the masks from the computer vision model for the region in (a). The center of the event is extracted as well as the maximum intensity along the event. (c) A 3D representation of the method to find the maximum intensity of a representative peak from (a). (d) The associated graph representation, where the marker size represents the maximum intensity along the events.}
    \label{DistributionBand}
\end{figure}

While HR-DIC provides information on an entire region, the automated segmentation of individual deformation events is desired to gain insight into the individual events. We used a computer vision model developed by Bean \textit{et al.} \cite{Bean2025} that utilized the existing YOLOV8 architecture to segment out the individual deformation events. The model was trained using HR-DIC measurements from a face-centered cubic (FCC) dataset. More information can be found elsewhere \cite{Bean2025}. The model identifies and segments each plastic deformation event, providing boxes and masks as depicted in Figure \ref{CV_HRDIC_small}(b, right). To utilize the masks identified by the computer vision algorithm, a custom script \cite{Bean2025} was developed to extract the maximum intensity along the events and their center positions. The centroid of each segmented region is computed as the arithmetic mean calculated by considering the segmentation region as a series of points in a 2D space. The maximum slip intensity value of each feature is determined through a multistage filtering process designed to robustly isolate the most significant peak within the region's intensity data. Initially, the intensity values are subjected to multiple rounds of statistical filtering based on dynamically computed thresholds of standard deviations from the mean intensity value. Each round refines the intensity values by removing outliers and focusing increasingly on the core of the distribution. The final maximum intensity value is then identified as the highest value within the remaining filtered and refined dataset. This method ensures that the maximum intensity value extracted is representative of the most prominent feature within the segmented region, while effectively minimizing the influence of background noise and outliers. A simplified representation of the developed procedure is shown in Figure \ref{DistributionBand}. For each deformation event, the maximum measured intensity, length, and center location are extracted. Using the extraction of these characteristics, we represent plastic deformation through the construction of a graph representation of the plastic deformation events for a given macroscopic deformation, and material composition and microstructure as illustrated in Figures \ref{CV_HRDIC_small}(c) and \ref{DistributionBand}(d). In these figures, each marker indicates a plastic deformation event, with its size and color representing the maximum intensity measured along this event. The lines or connections between events denote connectivity between the first 5 nearest neighboring events. The graph representation in Figure \ref{CV_HRDIC_small}(d) corresponds to the region within the white dashed box in Figure \ref{CV_HRDIC_small}(a).   

\begin{figure}
    \centering
    \includegraphics[width=1\textwidth]{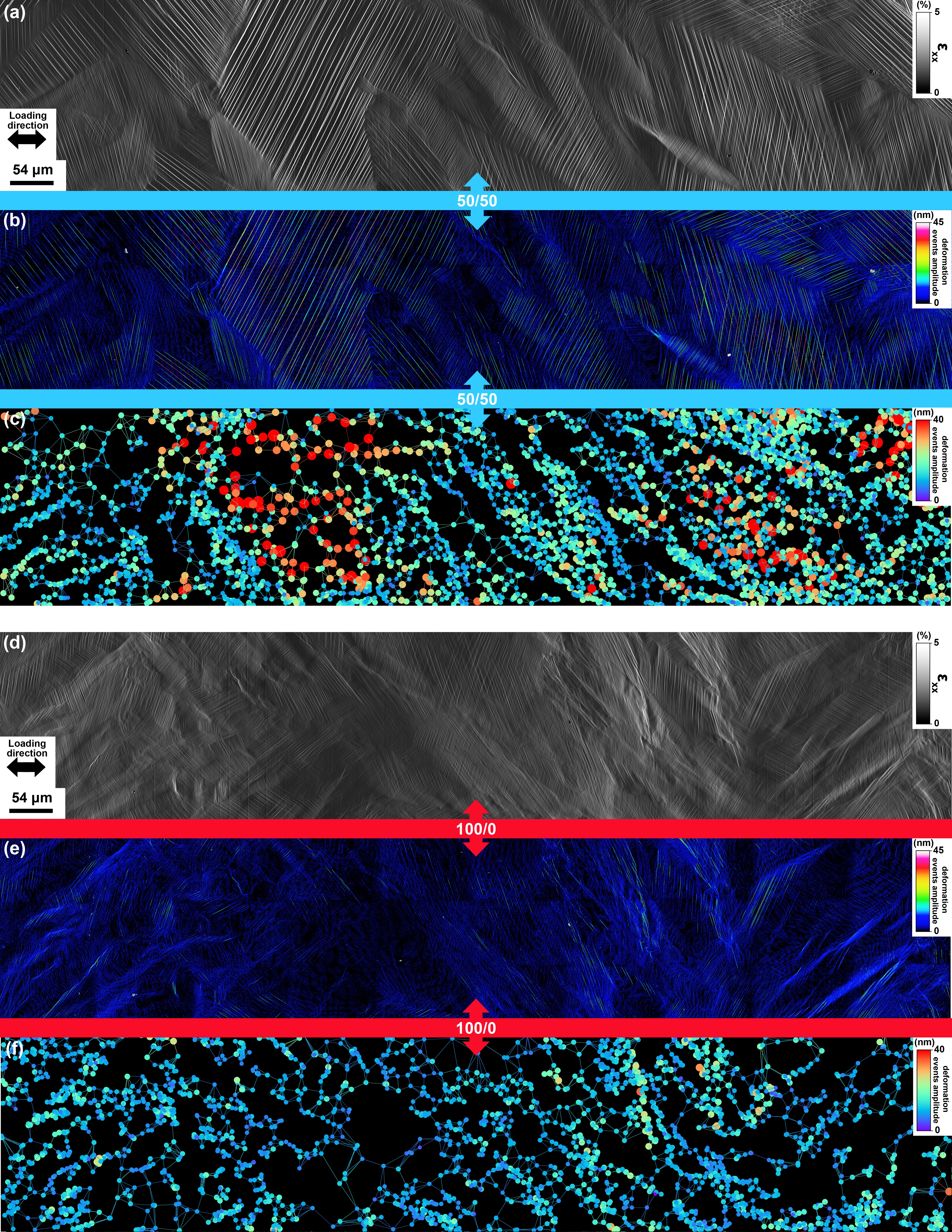}
    \caption{\textbf{Deformation event extraction for two regions of interest along the FGM gauge length} (a,d) $\epsilon_{xx}$ strain maps along the loading direction for two regions of interest with a ratio of 50/50 (a) and 100/0 (d) at applied macroscopic stress of 372.5 MPa. (b,e) Associated plastic deformation events intensity maps. (c,f) Associated graph representation, where marker size represents the maximum intensity along the events.}
    \label{HDICMethod_small}
\end{figure}

The described methodology was utilized to create a graph representation, for visualization and validation purposes,  of the plastic deformation events across the entire gauge length of the investigated graded material specimen. A massive dataset comprising 13,441,680,000 pixels with a resolution of $\unit{22}{\nano\meter}$ was collected during the tensile test, capturing the evolution of the plastic deformation events for the different investigated macroscopic plastic deformations. Automation facilitated the collection of 1,920 high-resolution images through the use of FEI Autoscript software \cite{autoscript4} coupled with a custom Python script for automated focusing, astigmatism correction, and grid imaging both before and after deformation. This script incorporates a registration technique to align images during the deformation steps to the initial images before deformation, ensuring post-deformation images are perfectly overlaid with their pre-deformation counterparts by adjusting the microscope stage position. The Python script has been made publicly available on GitHub at \url{https://github.com/cmbean2/Automated-SEM-Procedure}. Additionally, EBSD measurements were conducted over the entire specimen gauge length to inform the microstructure, using the parameters outlined in section 2.1. This allowed the creation of graph representations containing approximately 200,000 individual events for the entire specimen gauge length. Portions of the full graph representations across the gauge length of the graded stainless steel material are shown in Figure \ref{HDICMethod_small}(c,f) for the 50/50 (50\% 304L and 50\% 316L) and 100/0 (100\% 304L) composition ratios.

HR-DIC measurements can be performed at multiple deformation steps on the same region, enabling the analysis of the evolution of deformation events as a function of the applied macroscopic load or macroscopic plastic strain. Figure \ref{Interpolation}(a,b) illustrates an example region in the graded material subjected to applied macroscopic plastic strains of 0.07\% to 0.76\%. The corresponding graph representations of the regions are shown in Figure \ref{Interpolation}(b) overlapped with the $\epsilon_{xx}$ strain maps, highlighting the evolution of deformation events with increasing plastic strain, including the intensification of existing events and the occurrence of numerous new events. Only half of the maps have been overlapped with the graph representation to facilitate visualization. 

\subsubsection{Fatigue strength prediction from the maximum intensity of plastic deformation events}

\begin{figure}
    \centering
\includegraphics[width=1\textwidth]{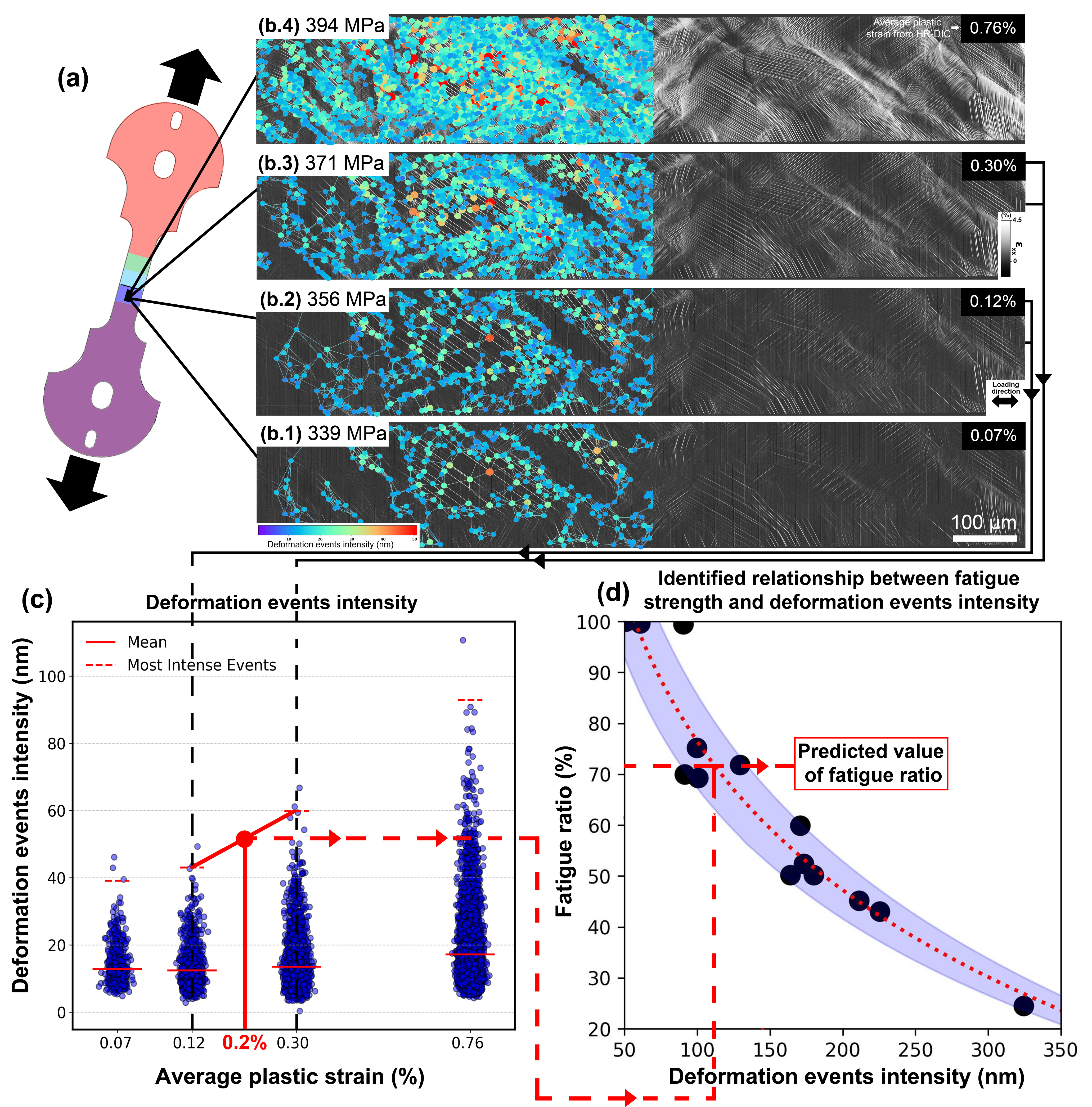}
    \caption{\textbf{Plastic strain ($\epsilon_{xx}$) and graph representation} (a) Specimen geometry with color indicating different chemical compositions. (b) Strain map ($\epsilon_{xx}$) of a region of interest (from the 25/75 region) in the graded material subjected to applied macroscopic plastic strains of 0.07\%, 0.12\%, 0.3\% and 0.76\%. Half of the maps have been overlapped with the graph representation of the segmented deformation events. (c) Statistical distribution of deformation event intensities as a function of applied macroscopic plastic strain. The most intense deformation events are identified for each applied macroscopic plastic strain, enabling the estimation of the average intensity of the most intense events at a macroscopic plastic strain of 0.2\% (yield strength). (d) This value is used to estimate the fatigue strength of the region based on the previously identified relationship detailed in Ref. \cite{science2022}.}
    \label{Interpolation}
\end{figure}

Identifying the most intense plastic localization events can enable the prediction of fatigue strength. This correlation was previously established in Ref. \cite{science2022} and is applied in this study to predict the fatigue strength across the different chemical compositions on the FGM, specifically as a function of the powder ratio of 316L/304L. The established relationship is illustrated in Figure \ref{Interpolation}(d) for a large set of metallic materials examined in Ref. \cite{science2022}. The fatigue ratio, as defined by Fleck \textit{et al.} \cite{FLECK1994365}, was obtained by conventionally measuring the yield strength and fatigue strength at $10^{9}$ cycles of several metallic materials \cite{science2022}. The fatigue strength were measured in the very high cycle fatigue regime using ultrasonic fatigue testing \cite{GEATHERS2022106672,science2022}. The maximum intensity of plastic deformation events was measured for each material using the identical HR-DIC method detailed in the present manuscript and at a macroscopic plastic strain of 0.2\% (yield strength of the material). The optimal fit to the data (red dashed line in Figure \ref{Interpolation}(d)) has been achieved using a logarithmic law, and the uncertainty is indicated by the blue shaded area in Figure \ref{Interpolation}(d), which represents an accuracy in the prediction of about 10\%. 

This relationship was established for a wide range of materials deformed to a plastic strain of 0.2\%. Achieving prediction of fatigue strength using this relationship requires extracting the characteristics of the deformation events corresponding to an average plastic strain of 0.2\%. While this is possible in single-composition materials, it is not easily feasible in graded materials. As a result, multiple deformation steps are necessary to ensure that all regions, corresponding to different powder ratios, reach at least a plastic strain of 0.2\%. However, precisely achieving 0.2\% strain in all regions during multiple deformation steps remains challenging. To address this, results can be interpolated between average macroscopic strains below and above 0.2\% for each region in the graded materials. The average plastic deformation is measured by averaging the HR-DIC measurement ($\epsilon_{xx}$, strain along the loading direction) across all of a region (powder ratio). 

As an example, Figure \ref{Interpolation}(c) reports the distributions of the maximum intensity for all the plastic deformation events extracted from the region shown in Figure \ref{Interpolation}(b) for a region of interest in the graded material, utilizing the computer vision algorithm and extraction procedure outlined in Figure \ref{DistributionBand}. From this distribution, the most intense events are identified, and the average maximum intensity of these events is calculated (see red horizontal dashed lines in Figure \ref{Interpolation}(c)). Therefore, a linear interpolation is done between these values to estimate the average maximum intensity of the most intense plastic deformation events at a 0.2\% macroscopic plastic strain (yield strength of the material). This final value allows the prediction of the fatigue ratio of the material. 

\subsubsection{Validation of the proposed method}


The proposed approach was tested on a functionally graded stainless steel specimen transitioning from 100\% 316L to 100\% 304L. In parallel, the fatigue strengths at $10^{9}$ cycles were measured for 100\% 316L and 100\% 304L materials, using the same printing parameters as the graded specimen. Eight conventional cylindrical fatigue specimens were extracted and machined from large AM blocks (20 mm $\times$ 60 mm rods) of 316L and 304L materials. The specimens were subjected to very high cycle fatigue (VHCF) testing, following the methodology described in Ref. \cite{science2022}. The same method was used to establish the relationship presented in Figure \ref{Interpolation}(d) \cite{science2022}. The fatigue strength was determined using four specimens for the 316L material and four specimens for the 304L material. The fatigue tests were conducted using an ultrasonic fatigue testing setup, specifically the 3R$\texttrademark$ VHCF MEG20TT apparatus. For the experiments, cylindrical samples were prepared with a gauge diameter of $\unit{5}{\milli\meter}$ and a length of $\unit{16}{\milli\meter}$, employing a low-stress grinding method for machining. Before the fatigue tests, the samples underwent mechanical polishing to achieve a finish with $\unit{1}{\micro\meter}$ diamond suspension. All testing was carried out at ambient temperature, adopting a 20 kHz frequency. To prevent self-heating, the "pulse-pause" technique was utilized, using 300 ms of activity followed by 700 ms of rest. The fatigue strength at $10^{9}$ cycles was estimated using the staircase approach, where the initial sample from each material batch was subjected to a low alternating stress. Intervals of 1\% power were used, corresponding to approximately 10 MPa. If no failure occurred after $10^{9}$ cycles, then the stress level was elevated. Subsequent samples were then tested at specific alternating stress levels to validate the fatigue strength obtained through the staircase method. The yield strength of both materials was determined through standard tensile testing of a specimen using monotonic deformation at quasi-static strain rate in the range of  $10^{-4}$~s$^{-1}$ to $10^{-3}$~s$^{-1}$.

\section{Results}



\subsection{Plastic deformation events characteristics at yield strength}

The graded specimen maintains a consistent cross-section along its gauge length, ensuring that the stress applied across all regions is uniform. However, the local strain experienced by each region varies. To determine the local applied plastic strain, we utilized the $\epsilon_{xx}$ strain maps obtained from HR-DIC measurements. Sections were analyzed within the strain maps along the gauge length, calculating their average $\epsilon_{xx}$ strain. For illustration, Figure \ref{HDICMethod_small}(a and d) shows the reduced region of the $\epsilon_{xx}$ strain maps for sections within the 50/50 and 100/0 powder ratio regions, respectively, under an applied macroscopic stress value of 372.5 MPa. Corresponding plastic deformation events intensity maps highlighting the plastic deformation events are presented in Figure \ref{HDICMethod_small}(b and e), respectively. Utilizing the method outlined in section 2.2.1, we obtained graph representations of the plastic deformation events from these maps, as shown in Figure \ref{HDICMethod_small}(c and f), respectively. Significant differences are observed in the characteristics of plastic deformation events between regions of different powder ratios. In the 50/50 ratio region, for instance, a noticeably lower number of plastic deformation events were detected compared to the 100/0 ratio region. However, the events in the 50/50 region exhibited, on average, significantly higher maximum intensity. Additionally, considerable variability was noted within the same region, with some areas showing a limited number of plastic deformation events and others displaying a multitude of high-intensity events.

\begin{figure}
    \centering
    \includegraphics[width=1\textwidth]{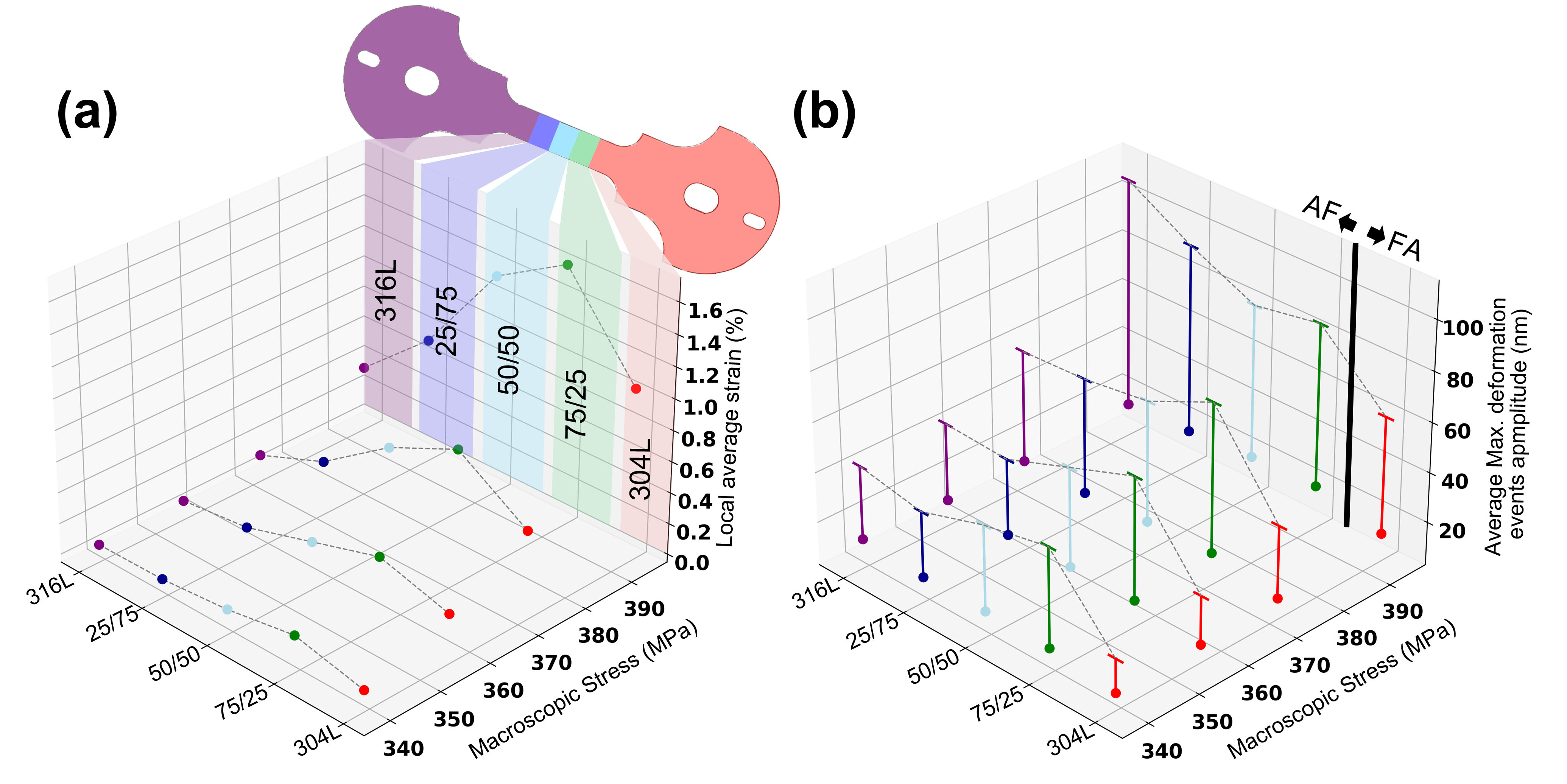}
    \caption{\textbf{Plastic strain and deformation event amplitude across the FGM gauge length} (a) Average $\epsilon_{xx}$ plastic strain for sections along the gauge length of the investigated graded specimen as a function of the applied macroscopic stresses. The color code corresponds to the different compositional ratios associated with the sections. (b) Average maximum plastic deformation events intensity for sections along the gauge length of the investigated graded specimen as a function of the applied macroscopic stresses.}
    \label{Strain}
\end{figure}

Figure \ref{Strain}(a) shows the average $\epsilon_{xx}$ plastic strain for all the sections along the gauge length of the graded specimen under different applied macroscopic stresses. The compositional ratios corresponding to the different sections that were obtained by chemical etching are indicated with the colors in Figure \ref{Strain}, following the color code defined in Figure \ref{FGM}(a). This visualization reveals that each compositional ratio region exhibits a distinct average $\epsilon_{xx}$ strain, indicating varied material responses depending on the ratio. Notably, the 316L section of the FGM tends to exhibit lower strain levels, whereas the 75/25 ratio material shows significant strain under the same applied macroscopic stress.

Furthermore, adhering to the methodologies detailed in sections 2.2.1 and 2.2.2, the average intensity of each section's most intense events is illustrated in Figure \ref{Strain}(b) across different applied macroscopic stresses. Interestingly, while the 75/25 ratio material demonstrates the highest strain, it does not display the highest plastic localization. Instead, the 0/100 ratio material exhibits the most intense plastic localization events, demonstrating a non-linear relationship between local $\epsilon_{xx}$ plastic strain and plastic localization.

\begin{figure}
    \centering
    \includegraphics[width=1\textwidth]{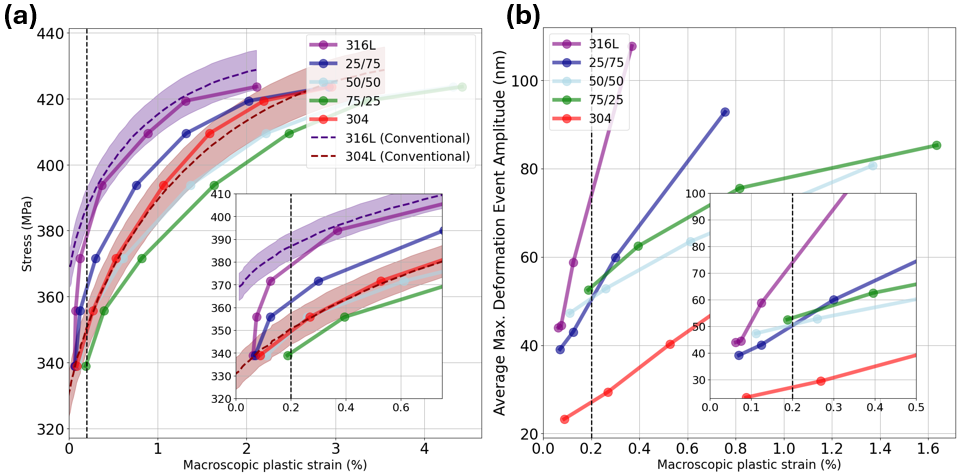}
    \caption{\textbf{Analysis by composition region across the gauge length} (a) Plastic strain-stress curves for each composition region along the gauge length of the graded material. (b) The average intensity of the most intense plastic deformation events as a function of the macroscopic plastic strain.}
    \label{Stress-SlipVSStrain}
\end{figure}

The average $\epsilon_{xx}$ plastic strain can be determined across all applied macroscopic stresses, enabling the acquisition of plastic strain-stress curves for each section. These curves, presented in Figure \ref{Stress-SlipVSStrain}(a), follow the established color code to differentiate between compositional ratios. Notably, the curves vary according to the compositional ratio, with the 25/75 ratio exhibiting a higher average $\epsilon_{xx}$ plastic strain compared to others. Similarly, Figure \ref{Stress-SlipVSStrain}(b) shows the average intensity of the most intense events for each region, plotted against the region's average $\epsilon_{xx}$ plastic strain. Using the data from Figures \ref{Stress-SlipVSStrain}(a) and (b), the macroscopic stress (yield strength) and the average intensity of the most intense events corresponding to an average $\epsilon_{xx}$ plastic strain of 0.2\% (material yield strength) can be accurately determined through interpolation, as explained in section 2.2.2. The interpolated yield strengths at a plastic strain offset of 0.2\% are shown in Figure \ref{Stress-SlipVSStrain}(a inset) for the different regions. Likewise, Figure \ref{Stress-SlipVSStrain}(b inset) shows the interpolated average intensities of the most intense plastic localization events at the plastic strain offset of 0.2\% for each region. 

\subsection{Fatigue strength prediction}

Utilizing the data on the average maximum intensity of the most intense plastic deformation events for each region, alongside the established relationship for predicting fatigue ratio depicted in Figure \ref{Interpolation}(d), we can determine the predicted fatigue ratio for each region. These predictions are detailed in Table \ref{Table}. Finally, leveraging the information on the yield strength corresponding to each region within the graded material, the fatigue strengths are obtained and presented in Table \ref{Table}. The 304L stainless steel portion of the gauge length exhibits the highest predicted fatigue strength with the proposed method. 

\subsection{Comparison with conventional mechanical data}

Table \ref{Table} shows the yield strength for AM 316L, 304L determined through conventional tensile testing on specimens with single composition as detailed in section 2.2.3. 
Concurrently, macroscopic fatigue specimens were obtained from large blocks of 316L, 304L and subjected to VHCF testing as outlined in section 2.2.3. These results are presented in Table \ref{Table} and labeled as "Conventional". The predictions derived from the high-throughput methodology described in this study from the graded material are included alongside in Table \ref{Table} for comparison purposes. For the 304L and 316L materials, similar results are observed between the conventional testing and proposed high-throughput method. A difference of less than 5\% is observed for the yield strengths, while a difference of less than 13\% is observed for the fatigue strengths. While the predicted fatigue strength of 316L is within 2\%, a higher difference of approximately 13\% is observed for 304L.


\begin{table}[h!]
    \centering
    \begin{tabular}{|c|c|c|c|c|c|}
        \hline
        Composition & Slip Intensity & Yield Strength & Yield Strength & Fatigue ratio & Fatigue ratio \\
        & (nm) & (MPa) & (MPa)  & (\%) & (\%) \\
        & (Accelerated) & (Accelerated) & (Conventional) & (Accelerated) & (Conventional)  \\
        \hline
        \cellcolor[HTML]{a672c2} 316L & 73.8 & 378.4 & 368.8 & 60.0 & 59.4 \\
        \cellcolor[HTML]{5c76cf} 75/25 & 50.28 & 362.5 & \cellcolor{gray} & 76.1 & \cellcolor{gray} \\
        \cellcolor[HTML]{7ee9ff} 50/50 & 50.62 & 348.9 &  \cellcolor{gray} & 75.8 & \cellcolor{gray} \\
        \cellcolor[HTML]{6ee37b} 25/75 & 53.14 & 339.9 & \cellcolor{gray} & 73.7 & \cellcolor{gray} \\
        \cellcolor[HTML]{ff7e7e} 304L & 27.09 & 349.3 & 355.7 & 101.8 & 85.6 \\
        \hline
    \end{tabular}
    \caption{Calculated values from the presented methodology and the comparative values from conventional testing.}
    \label{Table}
\end{table}

\section{Discussion}

This study presents an efficient, adaptable methodology for the high throughput prediction of fatigue properties using graded materials produced by additive manufacturing, here, focusing on stainless steels with varying ratios of nickel to chrome. Through our experiments, we have demonstrated the ability to evaluate the fatigue strength of five metallic materials with different compositions by analyzing the nanometer-scale plastic deformation events during monotonic deformation. We leverage inverse analysis between the statistics of plastic deformation events at the nanometer scale and fatigue strength. This approach avoids the time-consuming traditional mechanical testing. In the following sections, we discuss the advantages, future applications, and limitations of the proposed method, and highlight opportunities for further development.

\subsection{Rapid testing on FGMs specimens} 

The methodology described in this study enables testing of multiple chemical compositions using a single AM FGM specimen, significantly accelerating the evaluation of both yield and fatigue strength. Although this approach leads only to modest time savings for determining yield strength, which can already be conventionally measured rapidly, its impact on fatigue testing is substantial. Approximated time for each step of both the proposed high-throughput approach and VHCF testing are given in Table \ref{tab:time}. Under conventional conditions, even with already accelerated ultrasonic fatigue testing methods \cite{Cervellon2018}, evaluating the fatigue strength of a single alloy can span two to three months of continuous testing. In contrast, the measurements in our study took only two to three weeks in total, using a standard SEM with night-time operation only.

\begin{table}[]
    \centering
    \begin{tabular}{|c|c|c|}
        \hline
        & Proposed method & VHCF testing \\
        & (Accelerated) & (Conventional) \\
        \hline
        Samples preparation & (single specimen) $\approx$ 1.5 hours & (3 specimens tested per composition) $\approx$ 2 hours \\
        Mechanical testing & $\approx$ 1 hour &  (per composition) $\approx$ 277.8 hours\\
        Imaging & (per composition) $\approx$ 16 hours & / \\
        HR-DIC computation & (per composition) $\approx$ 24 hours* & / \\
        Computer vision & (per composition) $\approx$ 6 hours* & / \\
         \hline
    \end{tabular}
    \caption{Estimated time for each step of the proposed method and VHCF testing, based on average values from our tests on graded materials, 304L, and 316L. *This can be performed simultaneously with testing using a conventional GPU workstation equipped with two RTX 3080 graphics cards. }
    \label{tab:time}
\end{table}

Furthermore, if more advanced SEM technology, such as high-throughput multibeam systems were available \cite{Black2023}, the time dedicated to imaging could be shortened to just a few hours, offering an additional order-of-magnitude speedup. Given that five different compositions were assessed in a single FGM specimen, we achieved about a 20-fold reduction in testing duration. The exact extent of these time savings depends on the number of compositions per graded specimen, the availability of the microscope, and the specific equipment used. Under optimized conditions, i.e., with dedicated in-situ testing, high-speed SEM instrumentation \cite{Black2023} and GPU cluster, the entire workflow could be completed in a single day, enabling a potential 500-fold acceleration in fatigue-strength measurement.

\subsection{Deformation event analysis for fatigue strength evaluation}




The novel approach is particularly efficient for evaluating long-term properties, such as fatigue, that are resource-intensive to measure. Although a direct relationship between slip intensity and fatigue strength has been demonstrated \cite{science2022}, comparable relationships are yet to be identified for other critical long-term mechanical properties, such as creep, creep-fatigue and thermomechanical fatigue. This relies on identifying "elementary loading" to inform the long-term properties. In this study, a single cycle (monotonic loading) is considered and has been previously shown to provide insight into fatigue strength through the statistical measurement of deformation event intensity \cite{science2022}. This relationship also holds at higher temperatures, provided oxidation-assisted fatigue is not the dominant mechanism. As more comprehensive datasets of macroscopic properties and plasticity statistics are collected in the future, and relationships between plasticity during elementary loading and macroscopic properties are identified, the accuracy of the proposed method can be further improved, enabling its extension to other macroscopic properties.

Table \ref{Table} shows good agreement between the conventional testing and our described method for predicting both the yield strength and the fatigue strength within approximately 13\%. One source of variability is related to the limited study area. The statistically representative area of materials can play an important role in the estimation of mechanical properties. Materials produced by additive manufacturing are particularly concerned due to microstructural heterogeneities across various scales and over extended fields of view \cite{Wang2017}. As we investigate a limited field of view per composition, we may not capture the most intense deformation events (rare events), which are observed to control fatigue strength \cite{science2022,Prouteau2023}. In addition, the measurement is performed along a given surface of the specimen that may not be representative of the 3D bulk microstructure and therefore plasticity. Ultimately, this method could theoretically be extended to a continuous range of chemical compositions rather than a fixed number of compositions. As noted above, this approach involves a trade-off between the number of compositions within a single specimen and the statistical representativeness of a region. A larger region of interest provides a better estimate of the most intense events and therefore a more accurate fatigue strength prediction, but also a longer characterization time.

In the present study, measurements were obtained from only a single specimen and from a single surface, with each composition analyzed over only a few square millimeters. The proposed approach is based on the detection of the most intense events and therefore may overestimate the fatigue strength of materials. A comprehensive assessment of fatigue strength variability would require testing multiple specimens using the proposed method. Overall, the proposed method should be considered as a high-throughput approach for estimating fatigue strength to accelerate material design and enable the identification of promising materials and microstructures. 

The overall proposed framework can also be extended to various testing or characterization conditions, not just compositional variations. For example, the same strategy could be applied to a single material with testing temperature, grain size \cite{Machado2024}, or precipitation gradients \cite{Meng2019}. In each case, one specimen would suffice to explore a spectrum of properties, facilitating high-throughput screening and efficient alloy development. 

\subsection{Effect of microstructure variations across the FGM}

\begin{figure}
    \centering
    \includegraphics[width=1\textwidth]{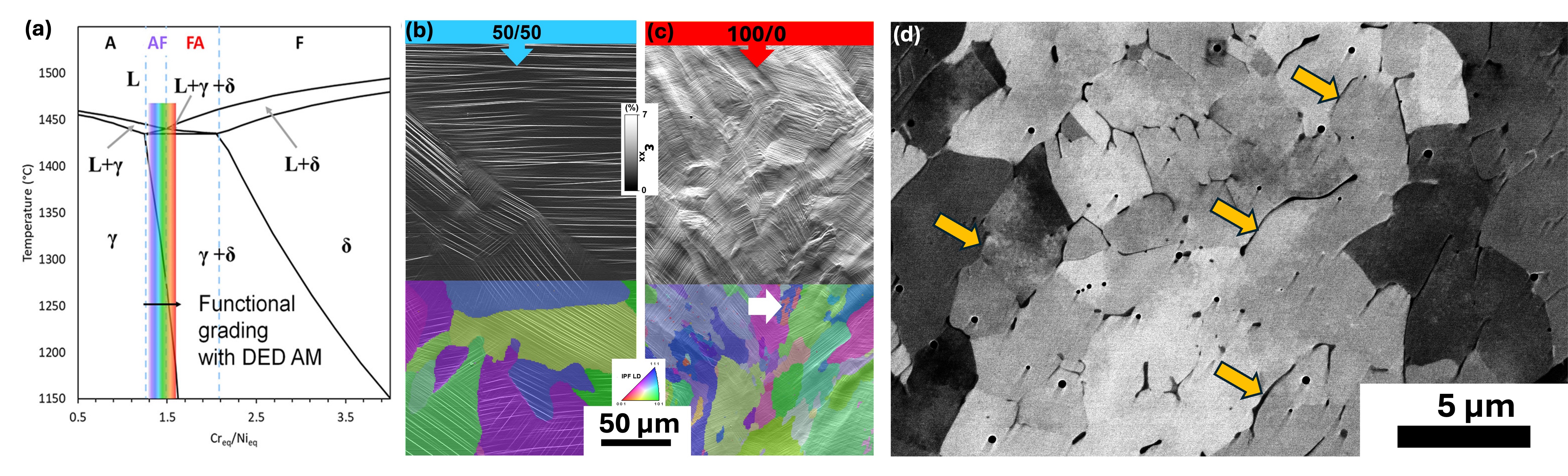}
    \textbf{Implication of solidification mode on microstructure development across the FGM.} \caption{(a) Pseudo-binary FeNi$_eq$Cr$_eq$ Shaeffler diagrams obtained from Thermocalc, adapted from \cite{NIE2023115714}  with permission of Elsevier. The solidification modes are highlighted with dashed lines, with the compositional ratio of the different regions on the graded material overlapped. The color code corresponds to the different regions in Figure \ref{FGM}. (b,c) $\epsilon_{xx}$ strain maps for a reduced region of interest for the composition of a ratio of 50/50 and 100/0 (304L) in the graded material. Associated inverse pole figure maps along the loading direction are overlapped. The white arrow indicates a twin. (d) Backscatter electron image in the 100/0 (304L) material showing $\delta$ ferrite particles pinning grain boundaries.}
    \label{AF}
\end{figure}

\begin{figure}
    \centering
    \includegraphics[width=1\textwidth]{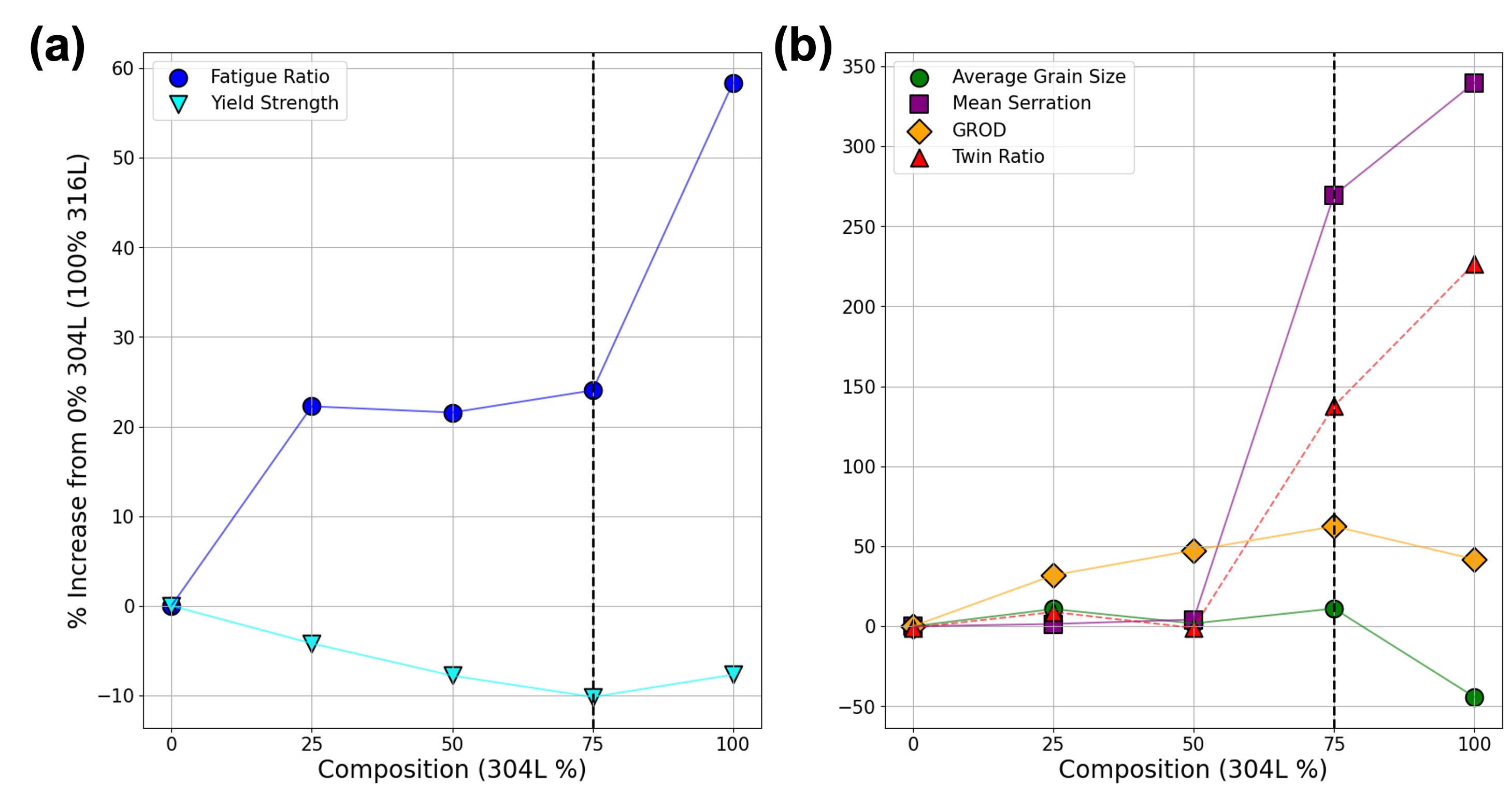}
    \caption{\textbf{Microstructural characteristics and macroscopic properties as a function of composition} (a) The normalized differences in percentage (compared to the 316L) for the predicted yield strength and fatigue ratios from the accelerated method. (b) The normalized differences (compared to the 316L) for the average grain size, the twin ratio, the mean grain boundary serration, and the grain reference orientation deviation (GROD) for the five different compositions across the sample.}
    \label{CompositionFigure}
\end{figure}

Interestingly, a significant increase in fatigue strength is observed between the AM 316L and 304L materials, occurring specifically in between the 25/75 and 304L regions. Shown in Figure \ref{AF}(b,c) are strain maps and the corresponding EBSD maps for regions within the 50/50 composition and the 100/0 (304L) compositions, respectively. Clear differences can be seen in the size of the grains and the nature of the plastic deformation events. The 50/50 region exhibits larger grains with long and highly localized slip bands. Conversely, the 100/0 (304L) region exhibits a finer grain structure with a high density of slip bands with low intensity. This transition in grain size is observed around the 75/25 composition, where the average grain size/diameter from the compositions with less than 75\% 304L is 170 microns; while the average grain size for the 304L structure is only 90 $\mu m$. A similar transition occurs in the amount of twin boundaries present. For the regions with less than 50\% 304L content, the twin ratio is close to zero, then at 75\% 304L, the ratio increases to 1.4\%, and reaches 2.3\% for the 304L region. Moreover, BSE imaging also revealed that in the 100\% 304L region, $\delta$ ferrite particles and cellular structures near the grain boundaries are present, whereas absent in 316L-rich regions. The development of these microstructural features is linked to a transition in solidification mode, shifting from austenite-to-ferrite (AF) to ferrite-to-austenite (FA) with increasing Cr/Ni ratio \cite{NIE2023115714}, as shown in Figure \ref{AF}(a). This change in solidification has a direct impact on the resulting microstructure, as detailed by Nie \textit{et al.} \cite{NIE2024120035}. Figure \ref{CompositionFigure}(b) summarizes the variation of several metrics characterizing the microstructural evolution as a function of composition, derived from EBSD measurements and compared to the fatigue ratio and yield strength in Figure \ref{CompositionFigure}(a). These include average grain size, twin ratio, mean grain boundary serration (calculated from the method described in ref. \cite{NIE2024120035}), and GROD. Grain boundary serrations are the result of dynamic GB pinning during solidification and the dissolution of the primary network of $\delta$ ferrite.  All these indicators use 316L as reference and their \% increase relative to 316L is reported. The transition in solidification mode is indicated by the dashed black line in Figure \ref{CompositionFigure}.

The drastic increase in the predicted (and conventional) fatigue ratio can be attributed to a significant reduction in slip intensity, as shown in Figure \ref{Strain}(b) (left and right of the dashed black line). As slip intensity decreases, the fatigue ratio increases \cite{science2022}. This reduction in slip intensity can be explained by various microstructural factors, primarily the decrease in grain size and the presence of twin boundaries, both of which reduce the mean free path of dislocations, thereby lowering slip length and intensity \cite{science2022}. Additionally, the presence of grain boundary serrations may contribute to the reduction in slip localization by facilitating slip activation. Moreover, fluctuations in GROD are also observed as a function of composition, which have been shown to influence slip localization \cite{Bean2025}. However, the GROD fluctuation remains of low amplitude, on the order of $1^{\circ}$.

Interestingly, in the region where the solidification mode is of AF type (316L, 25/75, 50/50, and 75/25), the predicted fatigue strength exhibits a slight increase with increasing Cr/Ni ratio. A decrease in yield strength is also observed, likely due to the influence of the Cr/Ni ratio on the critical resolved shear stress (CRSS) \cite{WAGNER2022117693}. Usually, an increase in yield strength is associated with more intense slip events, as observed in this study and reported elsewhere \cite{science2022, annurev:/content/journals/10.1146/annurev-matsci-080921-102621, refId0, CHAKRAVARTHY2010625, CLEVERINGA1999837}, leading to a reduction in the fatigue ratio \cite{science2022}.

\section{Conclusions}

The new ability to produce functionally graded AM materials provides unique opportunities to accelerate and reduce the time required to individually characterize different compositions. Using microscope automation, this study demonstrates the ability to collect high-resolution images over unprecedented amounts of the sample's surface. Then, by coupling this methodology with the existing HR-DIC methods and computer vision based segmentation, we were able to extract statistically significant information about the plastic deformation of each of the compositions across the sample. Ultimately, this enables the rapid and accurate estimation of fatigue strength for the diverse compositions using inverse relationships between local plasticity and macroscopic behavior. The proposed methodology offers many improvements over conventional testing methods that rely on sequential testing of different compositions. The method also shows potential to be used in various applications to investigate the effects of various parameters on the nanoscale deformation mechanisms and the corresponding macroscopic mechanical properties. \\


\noindent \textbf{Acknowledgments} \\

\noindent This work was carried out in the Materials Research Laboratory Central Research Facilities, University of Illinois. CB, MC and JCS are grateful for financial support from the the Defense Advanced Research Projects Agency (DARPA - HR001124C0394). CB, RLB, NV, EG, DA and JCS acknowledge the support of the National Science Foundation (NSF): DMR Grant \#2338346. Morad Behandish and Adrian Lew are acknowledged for their support and leadership. YN and MC acknowledge the NSF CAREER award \# DMR-2236640. Acknowledgment is extended to Valery Valle for providing the Heaviside-DIC code. \\

\noindent \textbf{CRediT authorship contribution statement} \\

\noindent \textbf{C.B.}: Conceptualization, Data curation, Formal analysis, Investigation, Methodology, Writing – original draft, Writing – review \& editing. \textbf{M.C.}: Conceptualization, Data curation, Formal analysis, Investigation, Methodology, Writing – original draft, Writing – review \& editing. \textbf{Y.N.}: Conceptualization, Data curation, Formal analysis, Investigation, Methodology, Writing – original draft, Writing – review \& editing. \textbf{R.L.B.}: Data curation, Formal analysis, Investigation, Methodology, Writing – original draft, Writing – review \& editing \textbf{N.V.}: Investigation, Methodology, Writing – original draft, Writing – review \& editing. \textbf{E.G.L.}: Data curation, Verification, Writing – review \& editing. \textbf{D.A.}: Investigation, Methodology, Writing – original draft, Writing – review \& editing. \textbf{M.A.C.}: Conceptualization, Funding acquisition, Methodology, Resources, Writing – original draft, Writing – review \& editing \textbf{J.C.S.}: Conceptualization, Funding acquisition, Methodology, Project administration, Resources, Supervision, Writing – original draft, Writing – review \& editing \\

\noindent \textbf{Declaration of Competing Interest} \\

\noindent The authors declare that they have no known competing financial interests or personal relationships that could have appeared to influence the work reported in this paper. \\

\clearpage

\bibliographystyle{unsrtnat}

\bibliography{Biblio}   

\clearpage

\end{document}